# Decision Support in the Context of a Complex Decision Situation

Teus H. Kappen[1], Mirko Noordegraaf[2], Wilton A. van Klei[1], Karel G.M. Moons[3], Cor J. Kalkman[1]

[1] Division of Anesthesiology, Intensive Care and Emergency Medicine

University Medical Center Utrecht

Utrecht, The Netherlands

[2] Utrecht School of Governance

Utrecht University

Utrecht, The Netherlands

[3] Julius Center for Health Sciences and Primary Care

University Medical Center Utrecht,

Utrecht, The Netherlands

**Abstract**

The aim of a clinical decision support tool is to reduce the complexity of clinical decisions. However, when decision support tools are poorly implemented they may actually confuse physicians and complicate clinical care. This paper argues that information from decision support tools is often removed from the clinical context of the targeted decisions. Physicians largely depend on clinical context to handle the complexity of their day-to-day decisions. Clinical context enables them to take into account all ambiguous information and patient preferences. Decision support tools that provide analytic information to physicians – without its context – may then complicate the decision process of physicians. It is likely that the joint forces of physicians and technology will produce better decisions than either of them exclusively: after all, they do have different ways of dealing with the complexity of a decision and are thus complementary. Therefore, the future challenges of decision support do not only reside in the optimization of the predictive value of the underlying models and algorithms, but equally in the effective communication of information and its context to doctors.

**Introduction**

Prediction models in health care estimate patient-specific risks of current or future conditions in new patients. By presenting patient-specific risk predictions to physicians, prediction models provide personalized decision support that aims to improve clinical decision making.[1,2] However, the evidence indicating that such decision support tools improve the quality of clinical decision making and patient outcome, remains highly inconsistent.[3–7] Current literature on clinical decision support suggests that adding therapeutic recommendations to the risk predictions will improve the effectiveness of the decision support tool.[6–8] The patient-specific risk prediction is merely a single factor in a highly complex decision making process. Consequently, the addition of a recommendation to the risk prediction may provide the clinical context that is required to make a personalized decision for an individual patient.[6,9]

In a recent series of studies, Kappen et al. added therapeutic recommendations to risk predictions of postoperative nausea and vomiting, which improved both the decision making of physicians and patient outcome.[9–11] Despite the improved effectiveness of the decision support tool, physicians did not comply with the recommendations in 34% of their patients, and many patients still suffered from nausea and vomiting after their surgical procedure. One explanation would be that the physicians took individual patient factors into account such as patient comorbidities or even patient preferences, and deliberately decided not to comply with the recommendations in 34% of their patients. An alternative explanation would be that the addition of recommendations did not provide the physicians with sufficient insight to value the recommendations for individual patients. The physicians could not decide whether the recommendation was either right or wrong, because in their view that depends entirely on the clinical context. Simply presenting more information to physicians is not the same as placing the information from the decision support tool within the clinical context – or *contextualizing* the decision support tool. In this article we will first argue that most current decision support tools are decontextualized. We then discuss how physicians rely on clinical context to handle the complexity of a clinical situation. Finally, we will argue that because of this complexity, decision support tools that contextualize predictions will be more effective in supporting physicians in decision making.

**Decision support**

'Clinical decision support' generally refers to electronic systems that aim to help physicians make better clinical decisions, which in turn should improve the outcome of their patients.[12] The concept here is similar to that of physician decision making: decision support uses individual patient and disease characteristics to provide physicians with patient-specific risk predictions or risk-based recommendations on treatment or diagnostic testing.[6,13] The predictions and recommendations may help physicians to analyze and manage the complexity of decision situations and reduce the cognitive workload of their decision making.[6,14] Current literature suggests that, for optimal effectiveness, decision support needs to blend in with the process of routine care and the presented predictions and recommendations should be embedded in the clinical context.[5,6,15–17] Contextualized decision support then minimizes the attentional resources that are required from physicians to interpret and use the predictions and recommendations in their decisions.

However, these claims seem to be largely unfounded, as decision support has failed to demonstrate that it improves patient outcome on a large scale.[5,18,19] From the theory of sociotechnical systems we know that the interaction between physicians and decision support tools will produce both 'designed' and 'unintended' changes in the existing decision making routines of physicians.[2,20–22] Such unintended consequences can paradoxically increase the cognitive workload of physicians or prevent the decision support from having a beneficial effect on patient outcome.[1,2] This does not mean that we should question the need for contextualized decision support. On the contrary, the unintended changes in decision routines and the inability to improve patient outcome indicate that many decision support tools are still decontextualized and do not necessarily manage complexity as well as they aim to.

**Decision complexity**

A decision becomes complex when the outcome of the decision is difficult to predict, despite the fact that one is reasonably informed about the individual components of the decision and their interrelations.[23] From this rather abstract definition we can establish that the level of complexity of a decision is not simply determined by the number of components and interrelations that are involved in a decision. Rather, the complexity depends on the (un)predictability of the effects of a decision from the components and interrelations that are known at the time of decision making. Consequently, in addition to the information that is available to the physicians, we also need to consider the information that the physicians are missing.

*The available information*

A basic decision on a medical intervention requires at least three pieces of clinical information: (1) the future risk of harm from the natural course of the disease (i.e. the prognosis); (2) the expected benefit from the intervention (i.e. the effectiveness); and (3) the nature and likelihood of harm from the intervention (side effects or complications). When these three informational components are exactly known for an individual patient the decision is relatively simple. It is a choice between two prognoses: either the risk of harm from the disease -- or the reduced risk of the disease after treatment, even after factoring the harms of the intervention. Since it is clear how the decision will affect outcome, the patient can weigh the two options according to his or her personal preferences and come to a decision.

   Such a clear-cut decision situation is merely hypothetical. Even for the simplest clinical decision the exact risks and benefits are not known for an individual patient and can only be approximated by probability estimations. There are three major methods for acquiring medical knowledge that a physician can use to estimate the probabilities for an individual patient: clinical research, pathophysiologic understanding, and clinical experience.[24,25] It is important to note that these methods do not necessarily have to estimate the probability in a quantitative way. To estimate the probabilities of risks and benefits for an individual patient, the knowledge that is acquired through each method has to be made patient-specific. This requires specific characteristics of the individual patient and the current disease state -- i.e. the clinical context -- such as the patient's demographics, lifestyle factors, results of diagnostic testing and comorbidity. The *complexity* of a clinical decision then becomes apparent. Many of these variables are interrelated, e.g. diagnostic testing depends on signs and symptoms, signs and symptoms may depend on comorbidity, etc. The greater the number of variables comprised in the clinical context, the greater number of interrelations will be involved. Further, each probability estimate for either a risk or an expected benefit is likely to be based on all three of above mentioned methods of knowledge acquisition, whereas all these methods use many of the same patient and disease variables. This greatly increases the number of possible interrelations in the decision. Even a basic medical decision is thus easily complicated by its clinical context.

*The missing information: uncertainty and ambiguity*

Notwithstanding the amount of information that is involved in the decision, it would still be possible to estimate the probabilities of risks and benefits for an individual patient when all the variables and interrelations are exactly known. It would be quite an undertaking for a physician -- although not insuperable -- whereas it would be very easy for a decision support tool. However, in most decision situations the variables and interrelations are not exactly known -- there is missing information. Most diseases do not manifest themselves through textbook signs and symptoms, patients do not always report all aspects of their lifestyle, and most diagnostic tests do not have a perfect predictive value. Accordingly, in many clinical decisions complexity truly complicates decision making.

How much the missing information increases the complexity of a decision depends on what part of the information is not available. When one knows that a particular variable or one of the interrelations is involved in a decision, but we do not know its exact value, the information may either be *uncertain* or *ambiguous*. Uncertainty means that we are informed about the value but it lacks precision[†] -- i.e. only the probability distribution of a variable or interrelation is known.[26–28] An imperfect diagnostic test is a good example of uncertainty, as we do know the probability of a diagnosis from the predictive value of the test, but we do not know whether in this instance the test is correct. Although uncertainty definitely adds to the complexity of a decision, with the right information it is still possible to calculate a probability estimate for the risks and benefits for an individual patient. The uncertainty in the variables and interrelations then results in uncertainty of the probability estimates for risks and benefits. A decision support tool is able to calculate the probability estimates with relative ease, through the use of sophisticated statistics and machine learning algorithms. Nonetheless, even under uncertainty, a rational patient would still be able to weigh the risks and benefits of treatment in a quantitative way and come to a decision according to his preferences.

---

[†] Precision in this instance does not refer to the statistical terminology of 'precision' and 'accuracy', as at the time of decision making one cannot distinguish between a random or systematic error for a single value.

Information is *ambiguous* when we understand the concept of the variable or the interrelation[‡] (we may even know the set of possible values), but we do not know the probability or the distribution of those values -- at least not in a quantitative way.[26–28] Ambiguity exists when a variable has multiple interpretations that coexist and may even oppose each other, or when its information is indistinct or ill-defined.[26,27] Ambiguous variables and interrelations are ample in clinical decision situations, e.g., atypical signs and symptoms, or inter-observer differences for a diagnostic test. As a result of the ambiguity in variables and interrelations, the expected risks and benefits can no longer be quantified. Ambiguity thus greatly decreases the predictability of the effects of a decision on patient outcome, which makes it much more difficult for patients to weigh risks and benefits.

Moreover, the preferences of patients -- or physicians -- are often ambiguous as well. For example, postoperative nausea and vomiting is very unpleasant for patients undergoing surgery and up to three different antiemetic drugs may need to be administered for effective prophylaxis,.[29] Although the risk of serious adverse drug events is probably very low (below one in one thousand), there is still a non-zero risk of a fatal cardiac arrhythmia or a serious infection of the surgical site. How does one value the risk of a fatal cardiac arrhythmia due to prophylactic treatment – even when its risk is one in one hundred thousand – against being nauseous and vomiting for 24 hours? Especially, when considering that the antiemetic drugs only reduce the risks of nausea and vomiting and cannot guarantee their prevention. Such ambiguity is more of an 'ethical' nature, in contrast to the more 'technical' ambiguity of the medical variables and interrelations.

---

[‡] Note that when a variable or interrelation is unknown, its concept is not known and there is no information, it does not directly contribute to the complexity, as such a variable contains information that is missing but not information that is available. Such an unknown factor is more likely to be a cause of ambiguity or uncertainty.

The ethical and technical ambiguity are also interrelated, as the difficulty for a patient to decide on such an ethical dilemma will depend on his or her ability to understand the risks and benefits that are presented by the physician. The complexity of a clinical decision is thus determined by the number of variables and interrelations that comprise the clinical context, the amount of uncertainty and ambiguity that is associated with them, and the ethical ambiguity of personal preferences. In the next sections, we will separately discuss the way that physicians and decision support tools handle complexity.

**Physicians and decision complexity**

Regardless of their complexity, clinical decisions will have to be made by patients and physicians. A good decision can only be made when all important elements -- variables, preferences, interrelations -- of the decision are considered, even when these elements are ambiguous. One strategy for physicians to handle the ambiguous elements in their decisions would be to somehow *interpret* and *weigh* the value of the elements. The estimated values for the ambiguous elements can then be used in the calculation of the probability estimates of risks and benefits. Nonetheless, the time and effort that would be required to weigh the value of all ambiguous elements and to analyze their interrelations with other decision elements would impose a far too great workload on a physician.[30,31]

Instead of relying on the slow and more analytic type of reasoning -- also known as Type 2 reasoning -- physicians rely on a faster and more intuitive type of reasoning -- known as Type 1 reasoning -- to deal with everyday decision complexity.[32–35] Type 1 reasoning uses experience-based strategies for problem-solving – or *heuristics* – that allow people to judge frequencies or possibilities in a fast and frugal way.[36] A physician will try to recognize the context of a complex decision, rather than try to understand the context and the complexity.[21,37,38] A recognized context becomes a frame of reference for all relevant information that the physician has learned to associate with the context, such as typical and atypical signs and symptoms, diagnostic testing, treatment options, etc. As the context is recognized in an intuitive, imprecise yet reliable way, it allows physicians to accept the complexity of a decision as it is – contextualization of the decision situation thus simplifies the decision itself. It then becomes much easier to interpret specific elements of the decision, to reason on risks and benefits and to communicate clinical information to other healthcare workers and patients.[39]

The 'fuzzy' recognition of complex patterns within the clinical context enables physicians to handle ambiguous information, which they may use in further analytic and reflective reasoning (Type 2 reasoning).[39,40] Physicians also acquire additional contextual information to understand how clinical information was produced, e.g. knowing who performed a diagnostic test may be informative of its reliability.[41] This production-contextual information may allow physicians to interpret otherwise ambiguous variables and weigh the value of those variables in their decisions. Contextualization of a decision situation thus provides a highly efficient heuristic for physicians to handle the complexity of clinical decisions while maintaining an acceptable workload.[39,42]

**Decision tools and decision complexity**

Decision support tools rely on the use of algorithms to provide risk-based predictions and recommendations. The algorithms may use any of the sources of medical knowledge to estimate the risks and benefits: clinical research, pathophysiologic understanding, and clinical experience. Today's computers are so powerful that decision support tools that must quickly handle lots of data are no longer limited by processing power. It thus seems intuitive that decision support tools are well equipped to model risks and benefits for complex decision situations, as long as there is sufficient data to do the necessary calculations. However, as decision support tools use algorithmic approaches to model the risks and benefits, they require quantifiable variables and interrelations as inputs. They thus do not easily handle ambiguous information. The greater the amount of ambiguity in decision situations, the greater the need for heuristics, which allow physicians to consider all available information, including the ambiguous elements.

**Decision tools and physician interaction**

The predictions and recommendations from decision support tools may still be very useful to physicians, even when there is ambiguity involved. For decision support tools to be useful in ambiguous decision situations, they have to effectively communicate their risk-based predictions in such way that physicians will be able to weigh the predictions in their decision making. When a decision support tool only presents quantitative risk predictions, it communicates decontextualized and analytic information to physicians, who may find it quite difficult to integrate in their more heuristic decision making process.[9]

The communication between a decision support tool and physicians is analogous to a native speaker who tries to communicate with a non-native speaker – as a decision tool can be considered 'native' in using probabilities and probability distributions. Even when the message is correctly phrased by the native speaker, it may require a lot of time and attention from the non-native speaker to understand the message. The message may even be incorrectly understood by the non-native speaker. Without a proper context the predictions are thus likely to increase the workload of physicians,[1,2] or may have an unintended negative effect on the quality of decisions – and may even result in a worse patient outcome.[20–22] This would make physicians fall back on their clinical experience and thus rely more on heuristics.

The difficulty to use the decontextualized predictions will prove greater for physicians when the decision complexity contains increased levels of ambiguity, as that would mean a greater gap between the predictions and the clinical context. Paradoxically, the more support a decision tool tries to provide in an ambiguous decision situation, the greater will be the risk of unintended changes in the physicians' routines, such as loss of situational awareness or degradation of clinical experience.[43] The negative consequences of a wrong decision on patient outcome are greater when a patient's risk of harm is higher. In other words, when the stakes of a decision are higher, the impact of a wrong decision will be greater, which requires a greater need to rely on physician heuristics.

**Contextual decision support**

From the previous sections we may conclude that the need for physicians to rely on heuristics – rather than to trust computerized decision support – will be greater when: (a) there is a lot of ambiguity, (b) the stakes of the decision are high, and (c) the decision support poses too much extra workload. The goals of contextualizing a decision support tool thus are: to make the tool acknowledge and 'factor in' ambiguity; to provide risk-based support to physicians; and to keep (or make) the workload acceptable. In the next sections, we discuss three ways to contextualize the predictions of decision support tools: the algorithms of the model may be changed to include more complexity; decision support tools may be more flexible to changes in clinical context; the communicative capabilities of the tool may be improved.

*The complexity model*

One way to contextualize a decision support tool is to improve the algorithms that the tool uses to calculate the predictions for the decision situation. The goal of the improvement would be to model all complexity. However, the challenge for the future will be to understand how we should deal with ambiguity in the algorithms of decision support tools. It is beyond the scope of this paper to discuss all available models and algorithms, or to speculate whether future artificial intelligence techniques will be able to handle ambiguity in a satisfactory way. Nevertheless, when ambiguity can be truly modeled by algorithms, the information is no longer ambiguous. A more promising approach is thus *disambiguation* of the decision, which will probably require specific research on the topic of the decision.

Alternatively, improving the data quality on which the algorithms are based would be another strategy, as it will improve the predictions of risks and benefits, despite the residual ambiguity. When we consider the example of postoperative nausea and vomiting, which was mentioned in the introduction, we would choose disambiguation of the decision. The risk predictions of which patients will become nauseous after surgery as well as the effectiveness of prophylactic treatment have been extensively quantified, albeit with a fair amount of uncertainty.[44–47] From the results of recent impact studies by Kappen et al., we started questioning whether there is an unquantified interrelation between the predicted risk and the effectiveness of treatment.[48] Further study may help to solve the ambiguous relation between predictions and treatment, or how to predict which patients will respond to treatment. Nevertheless, when a larger part of complexity is modeled by the algorithm, there is also an increased risk of unintended consequences, such as unwanted changes in the decision making routines of physicians.

*Flexible heuristic space*

A flexible heuristic space refers to the ratio between the use of heuristics and the use of predictive information in the decision making process. This concept is similar to that of the level of automation that is used in ergonomics.[43,49] The level of automation thus also depends on the stakes, the ambiguity, and the workload of the decision, as discussed in previous sections.

<u>Low ambiguity, low stakes.</u> When the level of ambiguity and the stakes are both low or when they are both high, the choice will be the easiest. For an unambiguous decision for which the stakes are low, human involvement may be restricted to the minimum and the process may be fully automated if at all possible.

High ambiguity, high stakes. High levels of ambiguity with very high stakes will require a fully autonomous physician, with the decision support tool functioning merely as an assisting instrument for information gathering and clinical reasoning. And perhaps – if properly designed and implemented -- it can also reduce the physician's workload.

Low ambiguity, high stakes. An unambiguous decision with high stakes poses a bigger challenge. As there is little ambiguity, it would make sense to let the decision support tool make the decisions to a great extent, with the physicians monitoring those decisions as the margin for error is small. Nonetheless, this situation is also prone to unintended effects of the decision support tool, as such a high level of automation may cause the physicians to lose situational awareness or degrade their clinical experience.[43] There are numerous examples from aviation where automation-related incidents could be traced back to loss of situational awareness by the crew.[50–52] It is thus not at all said that physicians will be able to keep the margin for error small when they are tasked with monitoring an almost fully automated process. As physicians will still be responsible for the final decision, it may prove more effective to lessen the level of automation. In fact, because the stakes are so high, the physicians may not even indulge such a high level of automation, especially when the decision support tool is perceived as counterproductive.[18,53,54]

High ambiguity, low stakes. A similar problem may occur when the decision is highly ambiguous, but when the stakes are low. The amount of ambiguity implies a greater need for heuristics. However, physicians are limited by their workload capacity.[35] Should this decision be the only problem that the physician faces at that moment, it would be highly efficient to simply let the physician make the decision, perhaps assisted by a computer. However, in a highly demanding environment, such as during surgery or treating a patient at the Intensive Care Unit, the physicians will not prioritize their attention to this relatively unimportant decision. In such an instance, it may be more efficient to let the decision support tool make a standardized recommendation, which directs the decision of the physician despite the presence of ambiguity.

These are four distinct situations. In clinical practice this will be more of a continuum. It is important to note that the level of automation does not automatically follow from the level of ambiguity and the stakes of the decision. The level of automation has to be chosen when designing the decision support tool. The ambiguity and the stakes of the decision may indicate what the risks of automating a decision will be, not whether the risks are acceptable.

Furthermore, the workload of the physicians in the decision situation at hand – from both the decision itself and other tasks and decisions – will determine how much time and attention can be additionally spent on using the decision support tool. In our example of postoperative nausea and vomiting, physicians had to make ambiguous decisions on preventive treatment in a high-workload environment for a problem that was not considered to be of utmost importance.[9] A higher level of automation – providing treatment recommendations that only required confirmation – was more effective than a lower level of automation, i.e. providing only risk predictions.[10,11] However, the effects on decision making and patient outcome were still modest when the decision support tool provided treatment recommendations. Going to the next higher level of automation – perhaps even fully automated decision support – may further increase the effectiveness of the tool, but may also increase the risk of unintended negative effects. The decision to further automate a decision support tool for postoperative nausea and vomiting may thus depend on the willingness of patients to accept the increased risks for a possible reduction in discomfort from nausea and vomiting after surgery.

Even within similar decision situations the level of ambiguity, the stakes and the workload may fluctuate. Ideally, the level of automation should be flexible, responding to these fluctuations. A future perspective might be that decision support tools do not only provide risk-dependent predictions or recommendations, but its level of automation also depends on the predicted risks and the workload of physicians. In our example of postoperative nausea and vomiting, the decision support tool might be able to estimate the workload of the physician at any time point during the anesthetic case. A flexible decision support tool would then provide its predictions and recommendations when there is a low workload during the case. Alternatively, a flexible tool would adjust its level of automation when the workload becomes too high.

*Communicative capabilities*

In current literature many factors are reported that improve the capabilities of a decision support tool to communicate context to physicians together with its calculated predictions. The hypothesis is that this will free up the mind of physicians to pay more attention to other complex activities that are not covered by the decision support tool.[55,56]

For example, the addition of actionable recommendations to risk predictions already aims to provide context on what a risk prediction may imply for treatment.[6,8,10,11] Alternatively, the reasoning behind the predictions may be provided, including details about the variables on which the model is based, and what predictors dominate in a particular patient.[6,15,18] This would facilitate physicians to better interpret the predictions and weigh them against their own clinical experience. Although there are several proven facilitators that improve the decision making of physicians, those interventions hardly incorporate ambiguity as a factor. As we argued in the section on 'Decision support', most decision support tools are still decontextualized. Ambiguity in clinical contexts causes information to be imprecise, thus true contextual decision support should not aim to provide precise information.[22,57] Decision support tools would be truly supportive when they communicate to physicians in the language of the physicians, instead of their own technical language of probabilities. Decision support should merely provide a description, which is written in a natural language and should try to convey meaning instead of facts.

When it is possible for a decision support tool to communicate ambiguous information to the physicians, the physicians would be able to enter a response to the decision support tool. As the decision support tool knows the quantitative meaning behind the otherwise ambiguous description that it provided to the physician, the tool may use the response for further calculations in self-learning algorithms. For example, the decision support tool may communicate that a patient has a high risk of postoperative nausea and vomiting and has no direct contraindications for preventive treatment. The physician may respond that he does not want to administer one of the possible drugs because the patient has several comorbidities. Although the decision support tool does not have to directly understand what the physician exactly means, the tool may learn in time to recognize patients which are considered to be too fragile by physicians to accept the risks of preventive treatment. When the decision support tool collected sufficient data on such patients it may either confirm or refute the risk and communicate that to the physicians. In this way, the algorithms of decision support may be able to include clinical intuition of physicians into their calculations.

*Patient preferences*

Similar communicative capacities of decision support tools may facilitate communication with patients, either directly with patients or through the physicians. Patients should be able and allowed to decide whether and how their diseases will be treated. It is a physician's job to provide the patients with the right decision situation and decision options that are appropriate for the clinical context. The more complex the clinical context and the decision situation, the more difficult it will be to provide the right information and context to patients. Without context the predictions and recommendations of a decision support tool may not mean much to patients. Providing them with contextual information that is communicated in natural languages may improve the patients' understanding of the decision situation. Such communication may be bidirectional. For example, rather than asking a patient to weigh a risk of one in one hundred thousand of a serious adverse event, to a risk of one in three of postoperative nausea and vomiting, it may be more useful to explore the preferences of patients and provide the context that they require. This may help patients to straighten out their thoughts on the decision, but also help physician to understand their patients.

    A decision support tool may use a patient's input to select appropriate information about the clinical context and the decision situation, but the tool may also weigh patient preferences in its recommendations. Moreover, flexibility of a decision support tool may also be determined by patient preferences. After all, it is very common in (non-automated) clinical practice that patients express their preferences and priorities to their physicians. In our example, a flexible decision support tool for postoperative nausea and vomiting would draw more attention from a physician when the patient indicated that preventing nausea and vomiting was highly important to him. In

other words, a flexible decision support tool would allow physicians to prioritize their time and attention according to patient preferences.

**Conclusion**

Clinical decision support aims to reduce the complexity of clinical decisions, yet if poorly implemented and used it may actually complicate care and confuse physicians. In this paper we argued that information from decision support tools is often removed from its clinical context. In contrast, physicians largely depend on clinical context to handle the complexity of their day-to-day decisions. Simply offering analytic information to physicians – without its context – may be similar to trying to get two people who speak different languages to have an elaborate conversation.

'Believers' of artificial intelligence argue that, since computers are much more capable to handle large amounts of information than humans, future elaborate decision support tools can replace physicians.[58–60] Simply typing 'Will computers replace doctors' into Google yields millions of hits discussing the subject.[§] However, information is not the same as knowledge.[61] Apart from the non-technical skills patients expect from their doctor, future *iDoctors* and *Doctor Algorithms* would need to be able to handle clinical context and decision complexity to a similar – or even better – degree as physicians. Obviously, computers are more capable to handle large number of variables and estimate levels of uncertainty, but physicians have the unique skill to use and interpret – albeit imperfectly – the entire clinical context in their decisions, taking into account all ambiguous information and patient preferences. Instead of arguing about who is ultimately better at clinical decision making, it may prove to be a more profitable strategy to use the best of both worlds and let doctors arrive at optimal patient-centered therapeutic decisions using tools on their computers for probability estimation and other applications requiring hardcore number crunching. To improve the quality of decision support tools, designers should focus not only on predictive accuracy, but also on optimizing the way in which such probabilities and therapeutic recommendations are communicated to and interpreted by doctors.

---

[§] At the time of writing, the search string 'Will computers replace doctors' without quotation marks yielded approximately 49,900,000 results when entered into Google. However, when the search string included quotation marks, it yielded only 1,410 results. Paradoxical to the conclusion of this paper, this is a nice example of artificial intelligence being able to handle context.